\documentclass{PoS}
\pdfoutput=1
\usepackage{graphicx}
\usepackage{amsmath}
\usepackage{amsfonts}
\usepackage{amssymb}
\usepackage[utf8]{inputenc}
\usepackage{cite}
\usepackage{breakurl}
\usepackage{physics}

\title{QED radiative corrections 
for Polarized Lepton-Proton Scattering}

\ShortTitle{QED radiative corrections 
for Polarized Lepton-Proton Scattering}

\author{\speaker{R.-D. Bucoveanu}\\
        PRISMA+ Cluster of Excellence, 
        Institut f\"{u}r Kernphysik, 
        Johannes Gutenberg-Universit\"{a}t, D-55099 Mainz, Germany\\
        E-mail: \email{rabucove@uni-mainz.de}}

\author{H. Spiesberger\\
        PRISMA+ Cluster of Excellence, 
        Institut f\"{u}r Physik, 
        Johannes Gutenberg-Universit\"{a}t, D-55099 Mainz, Germany\\
        E-mail: \email{spiesber@uni-mainz.de}}

\abstract{
We discuss QED radiative corrections at first and second order 
for the parity violating asymmetry $A_\text{PV}$ in elastic 
electron proton scattering. A measurement of $A_\text{PV}$ is 
planned at low energies and with high precision by the 
forthcoming P2 experiment at the new MESA facility in Mainz. 
Bremsstrahlung leads to a shift of the momentum transfer $Q^2$ 
which affects the determination of the weak charge of the 
proton.
}

\FullConference{
23rd International Spin Physics Symposium - SPIN2018 -\\
10-14 September, 2018\\
Ferrara, Italy}

\begin{document}

\section{Introduction}
\label{sec_introduction}

The upcoming P2 experiment at the MESA facility in Mainz plans 
to measure the weak mixing angle with a precision of $0.15 \%$ 
\cite{Becker:2018ggl} by determining the weak charge of the 
proton from a measurement of the parity violating asymmetry 
$A_{PV}$ of polarized electrons scattered off a proton target. 
The parity violating asymmetry is proportional to the momentum 
transfer $Q^2$ and is, therefore, affected by QED radiative 
corrections which lead to a shift of $Q^2$. In this work we 
discuss these radiative corrections at first and second order. 
A previous calculation of first order QED corrections to the 
parity violating asymmetry, can be found in 
\cite{Barkanova:2002,Aleksejevs:2007}.


\section{First-order QED corrections}

The leading order asymmetry between cross sections for incident 
electrons with positive and negative helicities, 
$\sigma^{(0)}_\pm$, at low momentum transfer, is given by
\begin{equation}
\label{eq:A_PV_0}
A_{PV}^{(0)}
= \frac{\sigma^{(0)}_+ - \sigma^{(0)}_-}%
       {\sigma^{(0)}_+ + \sigma^{(0)}_-}
= - \frac{G_F Q^2}{4\sqrt{2}\pi\alpha}\left[Q_W^p-F(Q^2)\right] 
\, ,
\end{equation}
where $Q_W^p$ is the weak charge of the proton and $F(Q^2)$ 
comprises form factors describing the proton structure. 
$F(Q^2)$ is small at the $Q^2$ values relevant for the P2 
experiment. Uncertainties due to $F(Q^2)$ will be known with 
sufficiently high precision from a combination of previous 
measurements and ancillary measurements at backward scattering 
angles planned at the P2 experiment. In Eq.~(\ref{eq:A_PV_0}) 
we see that the parity violating asymmetry is proportional to 
$Q^2$. Even though the QED corrections are parity conserving 
and do not affect the weak charge of the proton directly, 
they lead to a shift of the momentum transfer. If an experimental  
event-by-event determination of $Q^2$ is not possible, QED 
corrections have to be applied in the analysis to extract 
$Q_W^p$ from the measured asymmetry. Therefore, these corrections  
have to be calculated from theory with high precision. 

\begin{figure}[b!]\centering
\includegraphics[width=0.45\textwidth]{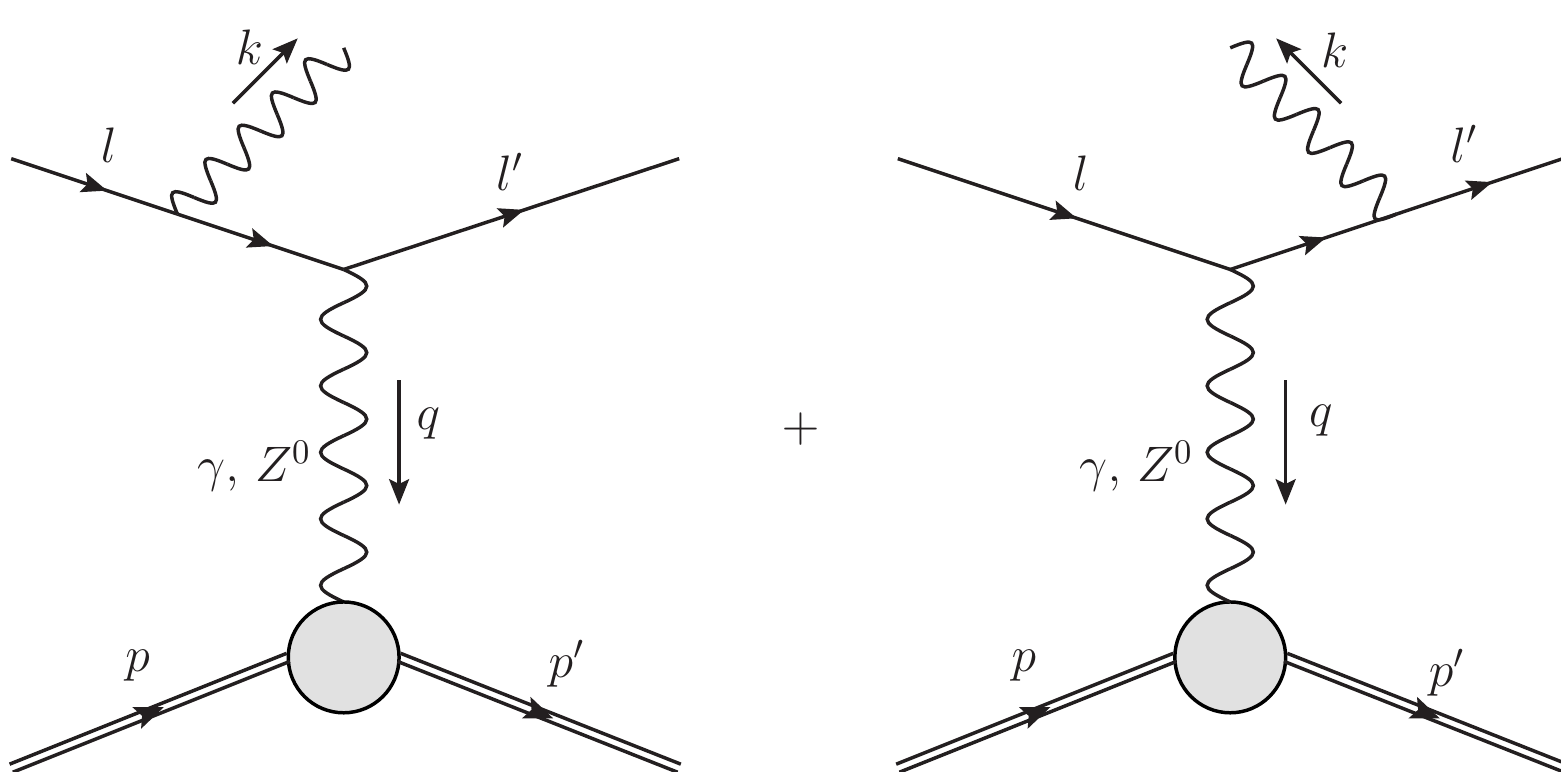}
\caption{Feynman diagrams describing first-order bremsstrahlung.} 
\label{Fig:1gamma_brems}
\end{figure}

\begin{figure}[t!]\centering
\includegraphics[width=0.55\textwidth]{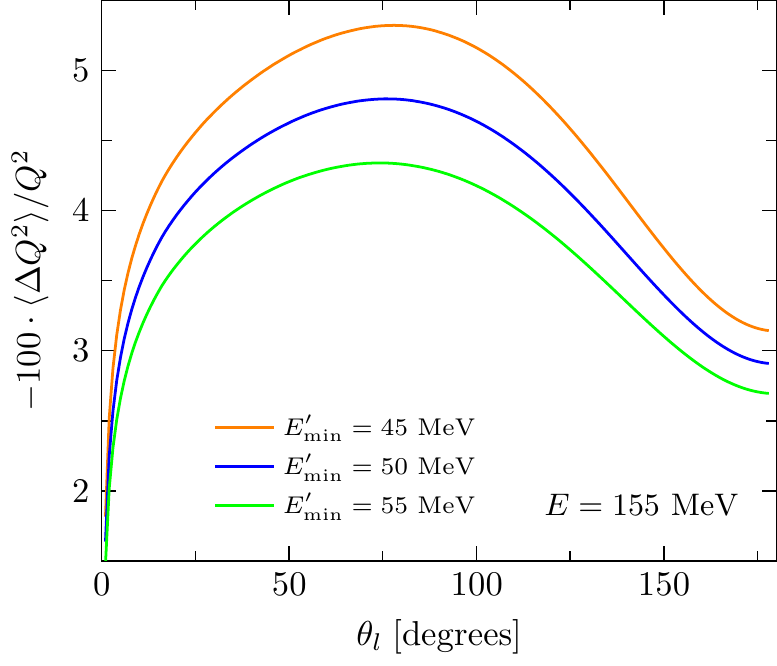}
\caption{ 
The average relative shift of the momentum transfer
due to hard photon radiation as a function of the 
electron scattering angle and for different minimum 
cuts for the energy of the observed scattered electron. 
The beam energy is $E=155$~MeV as expected for the 
Mainz P2 experiment. 
} 
\label{Fig:shift_Q2}
\end{figure}

From the measurement of the scattering angle one can only obtain 
information about what we can call leptonic momentum transfer 
(see \cite{Bucoveanu:2018soy}) defined as
\begin{equation}
\label{eq:leptonic_Q2}
Q^2_l = -(l-l')^2 \, ,
\end{equation}
where $l$ and $l'$ are the 4-momenta of the initial and final
electrons. However, taking into account bremsstrahlung effects
(see Fig.~\ref{Fig:1gamma_brems}), a photon with momentum $k$ will 
shift the momentum transfer to the true value given by
\begin{equation}
\label{eq:hadronic_Q2}
Q^2 = -(l-l'-k)^2 \, .
\end{equation}
We define an average shift of the momentum transfer by 
\begin{equation}
\label{eq:average_Q2}
\langle \Delta Q^2 \rangle 
= \frac{1}{\sigma ^{(0+1)}} \int \mathrm{d}^4 
\sigma _{1\gamma}^{(1)} \Delta Q^2 \, ,
\end{equation}
where $\sigma ^{(0+1)}$ is the total cross section including  
$\mathcal{O}(\alpha)$ corrections, while $\mathrm{d}^4 
\sigma _{1\gamma}^{(1)}$ is the differential cross section 
for the process with one additional photon radiated into the 
final state. For a more detailed description of these cross 
sections see Ref.~\cite{Bucoveanu:2018soy} and the discussion 
below. $\Delta Q^2=Q^2-Q^2_l$  is defined as the difference 
between the true $Q^2$ and the leptonic $Q_l^2$.
The integration in this expression can be performed numerically over 
the entire phase-space, since the product $\mathrm{d}^4 
\sigma _{1\gamma}^{(1)} \Delta Q^2$ is infra-red (IR) finite. 
Results for the kinematical conditions of the Mainz P2 experiment 
are shown in Fig.~\ref{Fig:shift_Q2}. We find that the average 
shift of the momentum transfer has a strong dependence on
kinematic variables. Its dependence on the scattering angle 
is seen in Fig.~\ref{Fig:shift_Q2}. We also find a strong 
dependence on the cut for the minimum value of the scattered 
electron's energy. This cut is related to the detector
acceptance of the experiment.

\begin{figure}[t!]
\centering
\includegraphics[width=0.75\textwidth]{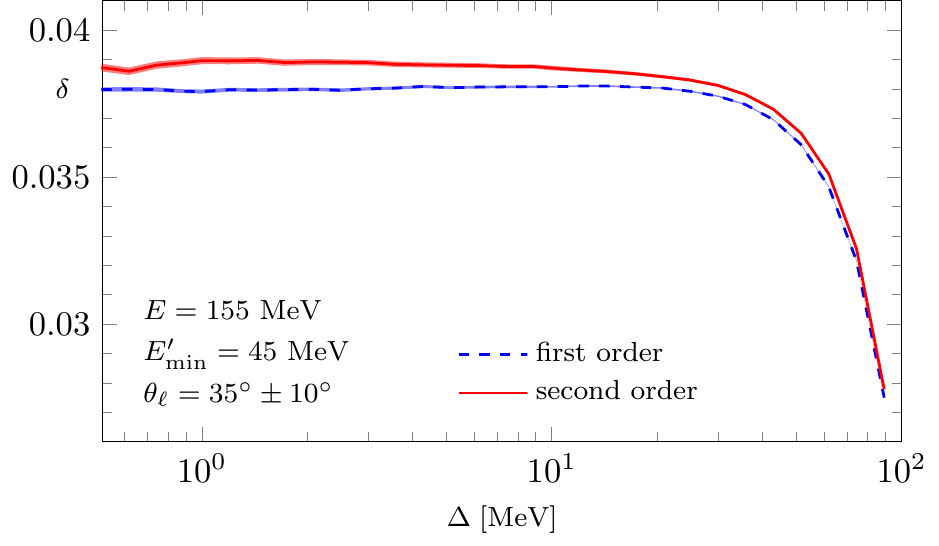}
\caption{ 
Test of the $\Delta$-independence of the complete correction 
factors when non-radiative contributions and hard-photon 
radiative effects are added, at first order $\delta = 
\sigma^{(0+1)} / \sigma^{(0)} -1$ and at second order $\delta = 
\sigma^{(0+1+2)} / \sigma^{(0)} -1$. Beam energy ($E = 155$~MeV), 
the range of the scattering angle ($25^\circ \leq \theta_l 
\leq 45^\circ$), and the minimum energy of the scattered electron 
($E_{\rm min}^\prime = 45$~MeV) is chosen for electron scattering 
at the Mainz P2 experiment. 
}
\label{Fig:total_cs}
\end{figure}

While a determination of the true momentum transfer is 
difficult or impossible, the dependence of the asymmetry 
on the electron scattering angle, $\theta_l$, is directly 
accessible in the experiment. We therefore study the effect 
of QED radiative corrections on the $\theta_l$-dependence 
of $A_{PV}$. The asymmetry including $\mathcal{O}(\alpha)$ 
QED corrections is defined as 
\begin{equation}
\label{eq_A_PV_1}
A_{PV}^{(0+1)}
= \frac{\sigma^{(0+1)} _+ - \sigma^{(0+1)} _-}{\sigma^{(0+1)} _+ + 
	\sigma^{(0+1)} _-}
= \frac{\sigma_P^{(0)}+\sigma_P^{(1)}}
{\sigma^{(0)}+\sigma^{(1)}} \, ,
\end{equation}
where the upper index $(n)$, $n = 0, 1,$ or $0+1$, denotes the 
order at which the cross section is evaluated, relative to the 
leading order. The polarization-dependent part of the cross 
section is labelled with a lower index $P$, i.e.\ 
$\sigma_P^{(n)} = \sigma^{(n)}_+ - \sigma^{(n)}_-$. 
The first order unpolarized cross section is given by 
(see \cite{Bucoveanu:2018soy})
\begin{equation}
\label{eq_sigma_unpol_1st}
\sigma^{(1)} 
= \sigma^{(1)}_{\text{non-rad}} + \sigma^{(1)}_{1h\gamma} \, ,
\end{equation}
where the non-radiative cross section, given by 
\begin{equation}
\sigma^{(1)}_{\text{non-rad}} 
= \int \mathrm{d}\sigma^{(0)} 
\Big[ \delta^{(1)}_{1-\text{loop}} + 
\delta^{(1)}_{1s\gamma}(\Delta) \Big],
\end{equation}
is IR finite and contains both 1-loop corrections and corrections 
due to one radiated soft photon. In this expression we have used 
correction factors relative to the Born-level cross section 
$\sigma^{(0)}$. The soft-photon part can be calculated 
analytically, integrating over the photon energy up to the 
cut-off $\Delta$, by using a soft-photon approximation 
as described in \cite{Bucoveanu:2018soy}. 
The cross section with a hard photon in the final state, 
i.e.\ with photon energies above the cut-off $\Delta$, 
is infrared finite and the phase space integration can be performed 
numerically. For one hard photon at tree level, we can write
\begin{equation}
\sigma^{(1)}_{1h\gamma} = \int _{E_\gamma > \Delta} 
\mathrm{d}^4\sigma^{(1)}_{1\gamma} \, .
\end{equation}
The sum of non-radiative and hard-photon contributions has 
to be independent of $\Delta$. However, when we use the 
soft-photon approximation to calculate the non-radiative 
contributions and due to numerical uncertainties we don't 
expect the result to be exactly $\Delta$-independent. We 
show the $\Delta$-dependence of the correction factor in  
Fig.~\ref{Fig:total_cs}. We find that the soft-photon 
approximation starts to distort the result at large values 
of $\Delta$. At very small $\Delta$, the numerical cancellation 
of $\Delta$-dependent terms lead to fluctuations and increasing 
numerical uncertainties. The fact that there is a wide plateau 
for which the numerical result for the cross section is 
independent of $\Delta$ proves that our approach leads to 
reliable predictions. Furthemore, we have checked that our 
calculation of the first order cross-section for scattering 
of unpolarized electrons agrees with \cite{Gramolin:2014pva} 
and \cite{Vanderhaeghen:2000ws}.

\begin{figure}[t!]\centering
\includegraphics[width=0.65\textwidth]{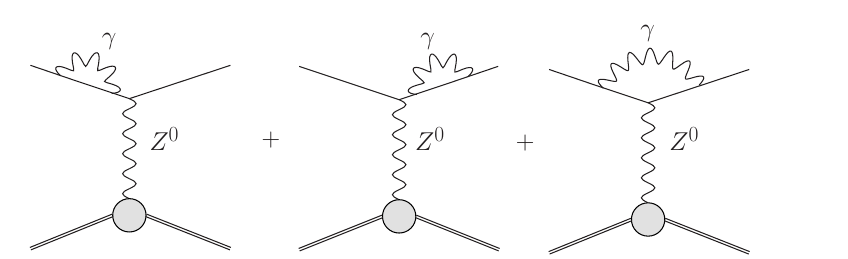}
\caption{One-loop photonic corrections to $Z_0$ exchange.} 
\label{Fig:Z0_1loop}
\end{figure}

In a similar way the polarized cross section can be separated into 
a non-radiative and a radiative part (see \cite{Bucoveanu:2019}) as
\begin{equation}
\label{eq_sigma_pol_1st}
\sigma^{(1)}_P
= \sigma^{(1)}_{\text{non-rad},\, P} + \sigma^{(1)}_{1h\gamma,\, P} \, ,
\end{equation}
where the non-radiative part is given, as for the unpolarized 
cross section, by the sum of QED 1-loop and one soft photon 
bremsstrahlung corrections. QED 1-loop corrections for one-photon 
exchange are the same as for the unpolarized cross section. 
The QED 1-loop corrections for $Z_0$ exchange are shown in 
Fig~\ref{Fig:Z0_1loop}. With on-shell renormalization, the 
self-energy diagrams vanish and the only diagram contributing 
in this case is the photonic vertex correction.

\begin{figure}[b!]\centering
\includegraphics[width=0.75\textwidth]
{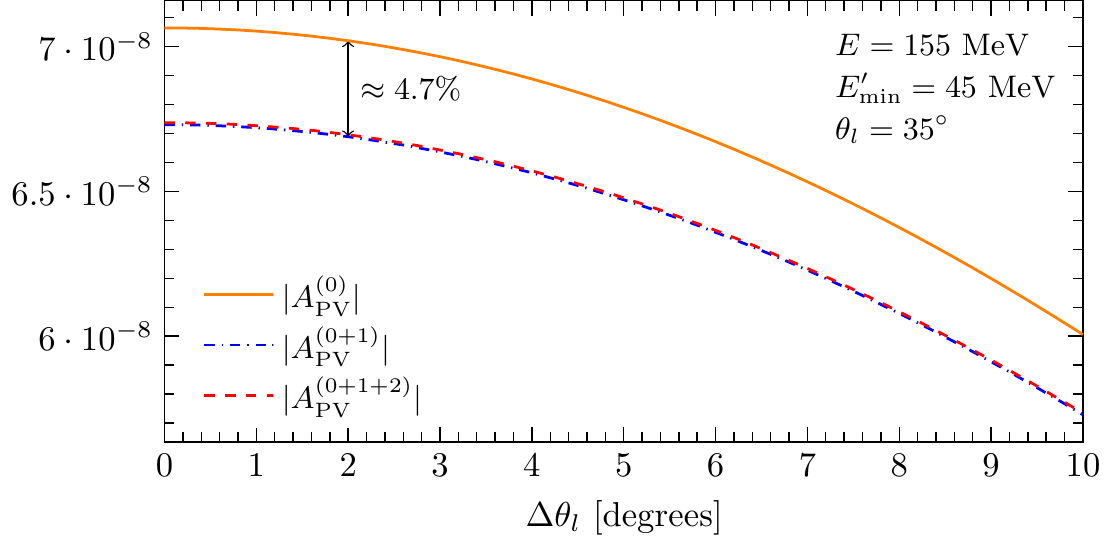}
\caption{ 
The parity violating asymmetry $A_{PV}$ at leading order 
and including $\mathcal{O}(\alpha)$ and $\mathcal{O}(\alpha^2)$ 
QED corrections. Kinematic variables are again chosen as 
relevant for the Mainz P2 experiment. The asymmetry is shown 
as a function of the scattering angle acceptance $\Delta\theta_l$, 
i.e.\ cross sections are integrated over the electron scattering 
angle in the range $35^\circ - \Delta\theta_l \leq \theta_l 
\leq 35^\circ + \Delta\theta_l$. 
} 
\label{Fig:total_asymm}
\end{figure}

Finally the hard-photon polarized cross section is defined as 
\begin{equation}
\sigma^{(1)}_{1h\gamma,\, P} = \int_{E_\gamma > \Delta} 
\mathrm{d}^4\sigma^{(1)}_{1\gamma,\, P} 
=
\int_{E_\gamma > \Delta} 
\left( \mathrm{d}^4\sigma^{(1)}_{1\gamma,\, +} - 
\mathrm{d}^4\sigma^{(1)}_{1\gamma,\, -} \right) \, .
\end{equation}
This contirubution to the cross section is IR finite and the 
integration can be performed numerically as in the case of 
the unpolarized cross section. The total asymmetry including 
$\mathcal{O}(\alpha)$ corrections is therefore also independent 
of the cut-off $\Delta$ and the difference between the leading 
order asymmetry and the asymmetry with $\mathcal{O}(\alpha)$ 
corrections is found to be of the same size as the relative 
shift of $Q^2$. This is demonstrated by the numerical results 
shown in Fig.~\ref{Fig:total_asymm}.


\section{Second-order QED corrections}

Since the first-order QED corrections lead to a large effect 
on the measured value of the parity violating asymmetry, it is 
important to include also $\mathcal{O}(\alpha^2)$ QED corrections, 
or to make sure they have a negligible effect.

The parity violating asymmetry including $\mathcal{O}(\alpha^2)$ 
corrections is written as  
\begin{equation}
\label{eq_A_PV_2}
A_{PV}^{(0+1+2)}
= \frac{\sigma^{(0+1+2)} _+ - \sigma^{(0+1+2)} _-}
{\sigma^{(0+1+2)} _+ + \sigma^{(0+1+2)} _-}
= \frac{\sigma_P^{(0)}+\sigma_P^{(1)}+\sigma_P^{(2)}}
{\sigma^{(0)}+\sigma^{(1)}+\sigma^{(2)}} \, ,
\end{equation}
where $\sigma^{(2)}$ is the $\mathcal{O}(\alpha^2)$ correction 
to the unpolarized cross section while $\sigma_P^{(2)}$ is 
the difference between the $\mathcal{O}(\alpha^2)$ corrections 
of the cross sections for positive and negative helicities of 
the initial electron. Following \cite{Bucoveanu:2018soy}, the 
unpolarized cross with second order corrections is written as
\begin{equation}
\label{eq:sigma_2nd}
\sigma^{(2)} 
= \sigma^{(2)}_{\text{non-rad}} 
+ \sigma^{(2)}_{1h\gamma} 
+ \sigma^{(2)}_{2h\gamma},
\end{equation}
where 
\begin{equation}
\sigma^{(2)}_{\text{non-rad}} 
= \sigma^{(2)}_{2-\text{loop}} 
+ \sigma^{(2)}_{1-\text{loop} + 1s\gamma} 
+ \sigma^{(2)}_{2s\gamma}
\, , \quad \quad 
\sigma^{(2)}_{1h\gamma} 
= \sigma^{(2)}_{1-\text{loop} + 1h\gamma} 
+ \sigma^{(2)}_{1s\gamma + 1h\gamma} \, , 
\end{equation}
and the labels indicate whether the different contributions 
are due to 1- or 2-loop diagrams or from 1- or 2- soft or 
hard photon radiation. The non-radiative part is rendered 
IR-finite by including loop diagrams: 
$\sigma^{(2)}_{\text{non-rad}}$ contains two-loop 
contributions and mixed soft-photon + one-loop parts, while 
$\sigma^{(2)}_{1h\gamma}$ contains one-loop corrections to the 
radiative process with one hard photon. An analytical expression 
for the 2-loop corrections can be found for example in 
\cite{Mastrolia:2003yz}. Using relative correction 
factors, the non-radiative part can be written as
\begin{equation}
\sigma^{(2)}_{\text{non-rad}} 
= \int \mathrm{d}\sigma^{(0)} 
\Big[ \delta^{(2)}_{2-\text{loop}} 
+ \delta^{(2)}_{1-\text{loop}+1s\gamma}(\Delta) 
+ \delta^{(2)}_{2s\gamma}(\Delta) \Big]\, .
\end{equation} 
Finally, the cross section for two hard photons is given by
\begin{equation}
\sigma^{(2)}_{2h\gamma} = \int_{E_\gamma , \; E'_\gamma > \Delta} 
\mathrm{d}^7\sigma^{(2)}_{2\gamma} \, .
\end{equation}
It is free of IR singularities and can be calculated 
numerically as described in \cite{Bucoveanu:2018soy}. The 
separate parts of Eq.~(\ref{eq:sigma_2nd}) depend on the cut-off 
$\Delta$, but the total result has to be independent of this
parameter. Figure~\ref{Fig:total_cs} shows that the cancellation 
of $\Delta$-dependent terms works also at second-order in a wide 
range of values for the IR cut-off $\Delta$.

As in the case of first order corrections, the second order QED 
corrections for the polarized cross section are defined in the 
same way as for unpolarized scattering: 
\begin{equation}
\label{eq:sigma_2nd_pol}
\sigma^{(2)}_P 
= \sigma^{(2)}_{\text{non-rad},\,P} 
+ \sigma^{(2)}_{1h\gamma,\,P} 
+ \sigma^{(2)}_{2h\gamma,\,P}.
\end{equation}
The definitions of the individual parts follow in the same way as 
for the unpolarized counter parts (see \cite{Bucoveanu:2019} for 
more details) and the total polarized cross section is IR finite 
and independent of the cut-off $\Delta$ on the radiated photons 
energies. A numerical calculation of the parity-violating 
asymmetry results in very small second-order corrections. 
This is shown in Fig.~\ref{Fig:total_asymm} where the curves 
for $A_{PV}$ at first and at second order are almost 
indistinguishable. In fact, radiative corrections for the 
asymmetry are due to the shift of the momentum transfer $Q^2$. 
This kinematical effect is already fully present if there is 
one radiated photon and second-order corrections contribute 
indeed only at the expected level with an additional factor of 
$\alpha/\pi$. Second-order QED corrections to the asymmetry 
have a negligibly small effect. 


\section{Conclusions}

In this work we discussed the effects of QED radiative corrections 
at first and second order to the parity violating asymmetry in 
elastic electron proton scattering. We found that, at first order, 
hard photon bremsstrahlung corrections lead to a shift of the 
momentum transfer $Q^2$ of approximately $5\%$ for P2 kinematics. 
Since $A_\text{PV}$ is proportional to $Q^2$ the difference 
between the leading order asymmetry and the asymmetry with first 
order QED corrections is also of this order of magnitude. 
Second-order QED corrections, however, contribute only very 
little. In addition to the QED corrections described above, 
also box graph contributions will have to be included in order 
to control the relation between the observed asymmetry and the 
weak charge of the proton at the required level of precision. 
A corresponding full calculation is underway.  


\section*{Acknowledgments}

This work is supported by the Deutsche Forschungsgemeinschaft 
(DFG) in the framework of the collaborative research center 
SFB1044 ``The Low-Energy Frontier of the Standard Model: From 
Quarks and Gluons to Hadrons and Nuclei''. 


\end{document}